\documentstyle[12pt]{article}
\def\thefootnote{\fnsymbol{footnote}}
\def\bea {\begin{eqnarray}}
\def\eea {\end{eqnarray}}
\def\be {\begin{equation}}
\def\ee {\end{equation}}
\def\ben{\begin{enumerate}}
\def\een{\end{enumerate}}
\def\bi{\begin{itemize}}
\def\ei{\end{itemize}}
\def\ie{{\it i.e.}\ }

\def\etal{{\it et al.}}

\def\F{{\cal F}}

\def\prl {Phys. Rev. Lett.\ }

\def\pr {Phys. Rev.\ }
\def\np {Nucl. Phys.\ }
\def\gA{g_{\mbox{\tiny A}}}
\def\gAeff{g_{\mbox{\tiny A,eff}}}
\def\qA{q_{\mbox{\tiny A}}}
\def\gL{g_{\mbox{\tiny L}}}
\def\gLeff{g_{\mbox{\tiny L,eff}}}
\def\qL{q_{\mbox{\tiny L}}}
\def\gLA{g_{\mbox{\tiny LA}}}
\def\gLAeff{g_{\mbox{\tiny LA,eff}}}

\def\gS{g_{\mbox{\tiny S}}}
\def\gSeff{g_{\mbox{\tiny S,eff}}}
\def\qS{q_{\mbox{\tiny S}}}

\def\gP{g_{\mbox{\tiny P}}}
\def\gPeff{g_{\mbox{\tiny P,eff}}}
\def\gPA{g_{\mbox{\tiny PA}}}
\def\gPAeff{g_{\mbox{\tiny PA,eff}}}

\begin{document}
\begin{titlepage}
\begin{center}
\vskip 1.0cm
{\large\bf Quenching of Spin Operators in the Calculation of }
\vskip 0.2cm
{\large\bf  Radiative Corrections for Nuclear Beta Decay}
\vskip 2cm
{\large I. S. Towner }\\
{\it AECL Research, Chalk River Laboratories, Chalk River}\\
{\it Ontario K0J 1J0, Canada}
\vskip 2cm
\today
\vskip 2cm

{\bf Abstract}
\end{center}

\begin{quotation}
Calculations of the axial-vector component to the radiative
correction for
superallowed Fermi $0^+ \rightarrow 0^+$
nuclear beta decay are here modified with
quenched rather than free-nucleon coupling
constants for the axial-vector and electromagnetic interactions
with nucleons.
The result increases the deduced value of $V_{ud}$ but does
not restore unitarity in the CKM matrix.
\end{quotation}

\end{titlepage}
\renewcommand{\thefootnote}{\#\arabic{footnote}}
\setcounter{footnote}{0}

Superallowed Fermi $0^+ \rightarrow 0^+$ nuclear beta decays \cite{HTKHS90}
provide both the best test of the Conserved Vector Current (CVC)
hypothesis in weak interactions and, together with the muon
lifetime, the most accurate value for the up-down quark-mixing
matrix element of the Cabibbo-Kobayashi-Maskawa (CKM) matrix,
$V_{ud}$.  Recent developments \cite{Wi90,Qu93}, however,
indicate a deterioration in the quality of the CVC test and a
lowering of the $V_{ud}$ value such that, with standard values
\cite{PDG92} of the other elements of the CKM matrix, the
unitarity test from the sum of the squares of the elements in the
first row fail to meet unity by more than twice the estimated
error.

Much of this deterioration is a consequence of the recent
improvements \cite{JR90,BAJR92,To92} in the calculation
of the nuclear-structure dependent part of the radiative
correction.
If the CVC hypothesis were correct, then the $\F t$ values
derived by correcting measured $ft$ values for the effects of
isospin-symmetry breaking and radiative corrections, should
be the same for all superallowed Fermi transitions
in all nuclei.  In the 1990 analysis \cite{HTKHS90} the $^{26m}$A$\ell$
data point had the lowest $\F t$ value; in particular it was lower
than the $^{14}$O data point.  However the revised
radiative correction calculation reverses this situation,
leaving the $^{14}$O data point with the smallest $\F t$ value.
Taken with the other seven precision data from
$^{26m}$A$\ell$ to $^{54}$Co the new analysis suggests a
$Z$-dependence
in the set of $\F t$ values, where $Z$ is the charge
number of the daughter nucleus in the beta decay.
Such a $Z$-dependence indicates
either an electromagnetic correction is still not accounted
for or that the CVC hypothesis is false.

Although we are discussing a purely vector interaction between
spin $0^+$ states, the axial-vector interaction does play a
role in the radiative corrections.  An axial-vector
interaction may flip the nucleon spin and then be followed
by an electromagnetic interaction that may flip it back again.
This axial contribution to the radiative correction was
considered by Marciano and Sirlin \cite{MS86}, who cast
the result into the following expression and estimated
its value:

\be
\frac{\alpha}{2\pi} [\, \ln (m_p / m_A ) + 2C \,]
= (0.12 \pm 0.18) \% .
\label{lnC}
\ee

\noindent   Here $\alpha$ is the fine
structure constant, $m_p$ the proton mass, $m_A$ a mass of order
1 GeV that provides a short-distance cut-off, and $C$ represents
the nonasymptotic long-range correction.  We write $C$ as

\be
C = C_{\rm Born} + C_{\rm NS} ,
\label{C12}
\ee

\noindent where $C_{\rm Born}$ refers to the Born graph in which the
axial-vector and electromagnetic interactions occur on the same
nucleon and $C_{\rm NS}$ is a nuclear-structure dependent
correction in which the interactions occur on different nucleons.

In the calculations \cite{JR90,BAJR92,To92} for $C$, the axial-vector
and electromagnetic vertices are evaluated with free-nucleon
coupling constants.  Yet there is ample evidence in nuclear physics
that coupling constants for spin-flip processes are quenched in
the nuclear medium \cite{BW83,To87}.
Thus the purpose of this Letter is to repeat
the calculations of \cite{To92} with quenched coupling
constants and investigate to what extent this ameliorates
the deterioration in the CVC test.

We assume that the axial-vector and electromagnetic vertices can
be described by on-shell form factors, even though in the
diagrams in question they are off-shell. Further we will use
nonrelativistic reductions and consider the form factors in the
zero-momentum limit and characterised by coupling constants
$g(k^2 \rightarrow 0)$.  These coupling constants can then
be equated with well-known coupling constants of nuclear
physics as deduced from electromagnetic $\gamma$-transitions
(and magnetic moments) and Gamow-Teller $\beta$-decay transitions
\cite{BW83}.

We follow the notation of \cite{To87,ASBH87} and write the
magnetic-moment operator as

\be
\mbox{\boldmath $\mu$}_{\rm eff}^{(I)}
= \gLeff ^{(I)} \mbox{\bf L}
+ \gSeff ^{(I)} \mbox{\bf S}
+ \gPeff ^{(I)} [Y_2,\mbox{\bf S}] ,
\label{mueff}
\ee

\noindent where $[Y_2,\mbox{\bf S}]$ represents a spherical harmonic
of rank 2, vector coupled to the spin operator, $\mbox{\bf S}$, to
form a spherical tensor of rank 1.  Here the superscript, $I$,
denotes the isospin structure: $I = 0$ being isoscalar
and $I = 1$ isovector.  The effective coupling constants are
written as

\be
g_{\rm eff} = g + \delta g ,
\label{geff}
\ee

\noindent where $g$ is the free-nucleon value and $\delta g$ a
nuclear-medium correction.  The free-nucleon values are:
$\gL^{(0)} =0.5$,
$\gL^{(1)} =0.5$,
$\gS^{(0)} =0.88$,
$\gS^{(1)} =4.706$,
$\gP^{(0)} =0.0$ and
$\gP^{(1)} =0.0$.
The nuclear-medium correction can also be expressed in terms of
a quenching factor

\be
q = g_{\rm eff} / g.
\label{quench}
\ee

\noindent  Calculations of the nuclear-medium correction $\delta g$
are given in \cite{To87,ASBH87} for closed-shell-plus (or minus)-one
nuclei $A = 5,15,17,39,41$.  They are based on corrections
to the single-particle wavefunction for these nuclei
being evaluated through to
second order in core polarisation, and on corrections
for meson-exchange currents and isobars.  Here we will use the values
from Table 26 of \cite{To87} and extrapolate or interpolate for
other mass values, $A$.
These values are in good agreement with the empirical values
deduced by Brown and Wildenthal \cite{BW83} in fits of shell-model
calculations to experimental data on magnetic moments and
M1 $\gamma$-transition rates in $sd$-shell nuclei.

Similarly we write the Gamow-Teller $\beta$-decay operator as

\be
(\mbox{\bf GT})_{\rm eff}
= \gLAeff  \mbox{\bf L}
+ \gAeff  \mbox{\boldmath $\sigma$}
+ \gPAeff [Y_2,\mbox{\boldmath $\sigma$}] ,
\label{GTeff}
\ee

\noindent and note the traditional use of the Pauli matrix
{\boldmath $\sigma$} rather than {\bf S}.  The free-nucleon values are
$\gLA = 0.0$,
$\gA = 1.26$ and
$\gPA = 0.0$.
Calculations of $\delta g$ can be found in \cite{To87,ASBH87}.
Here we will take values from Table 27 \footnote{There is one
typographical error in Table 27.  The entry for $\delta \gA$
for $A = 40$ $0d^{-1}_{3/2}$ should read $-0.255$.} of
\cite{To87} and again extrapolate or interpolate as required.
Here the results are not in such good accord with the
empirical values of Brown and Wildenthal \cite{BW83} obtained
in fits of shell-model calculations to experimental
Gamow-Teller $\beta$-decay rates.  For example, the quenching
factors $\qA$ from \cite{BW83} are 0.761, 0.737 and 0.727 for
A = 26, 34 and 38 respectively showing greater quenching
than the values we propose to use, as listed in Table 1, and
obtained from the calculations of \cite{To87,ASBH87}.
However the use of these stronger quenching factors for
$sd$-shell nuclei in the present analysis would not alter
significantly the conclusions to be drawn here.

\begin{table}[p]
\begin{center}
\caption{Quenching factors used in the present study  \label{tbl1}}
\vskip 2mm
\begin{tabular}{llllll}
\hline  \\[-3mm]
& \multicolumn {4}{c}{Electromagnetic} & Weak \\
\cline{2-5}   \\[-4mm]
& ~$\qL^{(0)}$
& ~$\qS^{(0)}$
& ~$\qL^{(1)}$
& ~$\qS^{(1)}$
& ~~$\qA$ \\[1mm]
\hline  \\[-3mm]
$A = 10$ & 1.042 & 0.897 & 1.173 & 0.927 & 0.878 \\
$A = 14$ & 1.044 & 0.873 & 1.201 & 0.934 & 0.858 \\
$A = 26$ & 1.023 & 0.869 & 1.146 & 0.877 & 0.835 \\
$A = 34$ & 1.026 & 0.850 & 1.155 & 0.870 & 0.812 \\
$A = 38$ & 1.028 & 0.840 & 1.159 & 0.866 & 0.801 \\
$A = 42$ & 1.010 & 0.862 & 1.133 & 0.866 & 0.824 \\
$A = 46$ & 1.010 & 0.857 & 1.137 & 0.862 & 0.818 \\
$A = 50$ & 1.011 & 0.853 & 1.141 & 0.857 & 0.812 \\
$A = 54$ & 1.011 & 0.849 & 1.145 & 0.854 & 0.807 \\[1mm]
\hline
\end{tabular}
\end{center}
\end{table}

\par In Table \ref{tbl1}, we list the quenching factors
to be used here.
For $p$-shell nuclei, they represent linear interpolations between
$A = 5$ and $A = 15$; for the $sd$-shell linear interpolations
between $A = 17$ and $A = 39$; while for the $pf$-shell
they are extrapolated from the $A = 41$ values using a scaling
factor of $(A/41)^{0.35}$ applied to $\delta g$.

For the case in which the axial-vector and electromagnetic interactions
occur at the same nucleon, the radiative correction is universal,
\ie the same for all nuclei, and has the value \cite{MS86,To92}

\be
C_{\rm Born}({\rm free}) = 3 \gA \gS^{(0)} I ,
\label{CBorn}
\ee

\noindent where $I$ is a loop integral.  With the replacements
$\gA \rightarrow \qA \gA$ and
$\gS^{(0)} \rightarrow \qS^{(0)} \gS^{(0)}$
the universality is now broken, and the contribution from $1$-body
graphs is written

\bea
C_{\rm Born}
& = & 3 \gA \gS^{(0)} I + (\qA \qS^{(0)} -1) \,
3 \gA \gS^{(0)} I
\nonumber \\
& = & C_{\rm Born}({\rm free})  +
(\qA \qS^{(0)} -1) \,  C_{\rm Born}({\rm free}) ,
\label{C1q}
\eea

\noindent where the second term becomes part of the nuclear-structure
dependence of the radiative correction.  For
$C_{\rm Born}({\rm free})$, we use
the value $0.881 \pm 0.030$ \cite{To92}.

\begin{table}[p]
\begin{center}
\caption{Revised values for the nuclear-structure part of $C$ obtained
through the introduction of quenching factors \label{tbl2}}
\vskip 2mm
\begin{tabular}{lrrc}
\hline  \\[-3mm]
& Unquenched & \multicolumn{2}{c}{Quenched} \\
\cline{3-4}  \\[-3mm]
& \multicolumn{1}{c}{$C_{\rm NS}$}
& \multicolumn{1}{c}{$C_{\rm NS}$}
& $(\qA \qS^{(0)} -1) C_{\rm Born}({\rm free})$
\\[1mm]
\hline  \\[-3mm]
$A = 10$ & $-1.67 \pm 0.20$ & $-1.35 \pm 0.16$ & $-0.19$ \\
$A = 14$ & $-1.15 \pm 0.30$ & $-0.88 \pm 0.23$ & $-0.22$ \\
$A = 26$ & $ 0.25 \pm 0.05$ & $ 0.20 \pm 0.04$ & $-0.24$ \\
$A = 34$ & $-0.17 \pm 0.06$ & $-0.13 \pm 0.05$ & $-0.27$ \\
$A = 38$ & $-0.10 \pm 0.10$ & $-0.09 \pm 0.09$ & $-0.29$ \\
$A = 42$ & $ 0.50 \pm 0.10$ & $ 0.40 \pm 0.07$ & $-0.26$ \\
$A = 46$ & $ 0.16 \pm 0.03$ & $ 0.14 \pm 0.03$ & $-0.26$ \\
$A = 50$ & $ 0.16 \pm 0.03$ & $ 0.14 \pm 0.03$ & $-0.27$ \\
$A = 54$ & $ 0.20 \pm 0.03$ & $ 0.17 \pm 0.03$ & $-0.28$ \\[1mm]
\hline
\end{tabular}
\end{center}
\end{table}

\par  For the two-nucleon graphs the operator is complicated
and comprises 12 terms as listed in Table 2 of \cite{To92}.  Different
terms originate in different pieces of the electromagnetic couplings.
Terms 1 and 2 are proportional to the isoscalar spin coupling,
$\gS^{(0)}$; terms 3 and 4 proportional to $\gS^{(1)}$;
terms 5, 6, 9 and 10 proportional to $\gL^{(0)}$; and terms 7, 8,
11 and 12 proportional to $\gL^{(1)}$.  The prescription then
is to modify the $2$-body operator by its appropriate electromagnetic
quenching factor and by the weak
quenching factor, $\qA$.  Numerical results are given
in Table \ref{tbl2}.

The data base for superallowed $\beta$-transitions produced by
the Chalk River group \cite{HTKHS90} in 1990 is here updated to
include four new lifetimes \cite{Ko93} and four new $Q$-values
\cite{Ko93,Ki91,Ba94}.  The methodology for handling the data and the
theoretical corrections remains the same as that used in
\cite{HTKHS90,To92} except for the introduction of quenching
factors in the radiative corrections.  The corrected $ft$-values,
$\F t$, for the eight precision data cases are fitted by a
one-parameter function

\be
\F t = \F t(0) = {\rm constant}
\label{Ft0}
\ee

\noindent or a two-parameter function

\be
\F t = \F t(0)\,[1+a_1Z]
\label{Ft1}
\ee

\begin{table}[p]
\begin{center}
\caption{Fitted values of $\F t(0)$ and $a_1$
for the eight precision superallowed Fermi $\beta$-decay data,
and the deduced values of $V_{ud}$ and the unitarity sum
\label{tbl3}}
\vskip 2mm
\begin{tabular}{ccc}
\hline  \\[-3mm]
& Unquenched & Quenched \\
& $C_{\rm NS}$
& $C_{\rm NS}$ \\[1mm]
\hline  \\[-3mm]
1-parameter fit & & \\
\cline{1-1} \\[-4mm]
$\F t(0)$ & $3075.0 \pm 3.5$ s & $3073.0 \pm 3.3$ s \\
$\chi^2/N$ & 1.91 & 1.40 \\
$V_{ud}$ & 0.9733 $\pm$ 0.0007 & 0.9736 $\pm$ 0.0007 \\
$V_{ud}^2+V_{us}^2+V_{ub}^2$
& 0.9959 $\pm$ 0.0016 & 0.9965 $\pm$ 0.0015 \\[2mm]
2-parameter fit & & \\
\cline{1-1} \\[-4mm]
$\F t(0)$ & 3067.4 $\pm$ 3.5 s & 3067.1 $\pm$ 3.3 s \\
$a_1$ & (1.4 $\pm$ 0.5) $\times 10^{-4}$ & (1.1 $\pm$ 0.5) $\times 10^{-4}$ \\
$\chi^2/N$ & 1.17 & 0.94 \\
$V_{ud}$ & 0.9745 $\pm$ 0.0007 & 0.9745 $\pm$ 0.0007 \\
$V_{ud}^2+V_{us}^2+V_{ub}^2$
& 0.9982 $\pm$ 0.0016 & 0.9983 $\pm$ 0.0015 \\[1mm]
\hline
\end{tabular}
\end{center}
\end{table}

\par The results are given in Table \ref{tbl3}, both with and without
the quenching factors.  In both cases the introduction of quenching
factors improves the fit (reduces the $\chi^2$).  In the 1-parameter
fit there is a reduction  of 2.0s in $\F t(0)$ and a concomitant
increase in $V_{ud}$, but not enough to restore unitarity.  In the
2-parameter fit, the intercept $\F t(0)$ is essentially unchanged
and hence there is little change in $V_{ud}$, but the slope
$a_1$ is reduced.  This indicates that the quenching factors are
responsible for about 20\% of the putative $Z$-dependence in the
current data base.  Thus, in spite of the considerations given here,
the deterioration in the CVC test in the precision superallowed
Fermi decay data persists.

The author acknowledges a conversation at the WEIN92 conference
with Prof. Yu. V. Gaponov, who insisted a calculation such as this
should be performed.

\end{document}